\documentclass[lettersize,journal]{IEEEtran}
\usepackage{amsmath,amsfonts}
\usepackage{algorithmic}
\usepackage{algorithm}
\usepackage{array}
\usepackage[caption=false,font=normalsize,labelfont=sf,textfont=sf]{subfig}
\usepackage{textcomp}
\usepackage{url}
\usepackage{verbatim}
\usepackage{graphicx}
\usepackage{cite}
\usepackage{pgfplots} 

\usepackage{tabularx}
\usepackage{array}

\usepackage{subcaption}
\usepackage{graphicx} 

\usepackage{svg}
\usepackage{colortbl}
\usepackage{graphicx}
\usepackage{multirow}
\usepackage[normalem]{ulem}
\usepackage{booktabs}
\useunder{\uline}{\ul}{}
\usepackage{mathtools}
\usepackage{hyperref}
\usepackage{makecell}
\usepackage{tikz}
\usepackage{listings}
\usepackage{tablefootnote}
\usepackage{amssymb}
\usepackage{pifont}
\usepackage{dblfloatfix}
\usepackage{array}
\usepackage{float}
\usepackage{wasysym}

\title{TelecomRAG: Taming Telecom Standards with Retrieval Augmented Generation and LLMs}

\author{Girma M. Yilma,~\IEEEmembership{Member,~IEEE,} Jose A. Ayala-Romero, Andres Garcia-Saavedra,  and Xavier Costa-Perez,~\IEEEmembership{Senior Member,~IEEE}}

\begin{document}

\maketitle

\begin{abstract}
Large Language Models (LLMs) have immense potential to transform the telecommunications industry. They could help professionals understand complex standards, generate code, and accelerate development. However, traditional LLMs struggle with the precision and source verification essential for telecom work. To address this, specialized LLM-based solutions tailored to telecommunication standards are needed. Retrieval-augmented generation (RAG) offers a way to create precise, fact-based answers. This paper proposes TelecomRAG, a framework for a Telecommunication Standards Assistant that provides accurate, detailed, and verifiable responses. Our implementation, using a knowledge base built from 3GPP Release 16 and Release 18 specification documents, demonstrates how this assistant surpasses generic LLMs, offering superior accuracy, technical depth, and verifiability, and thus significant value to the telecommunications field.


\end{abstract}


    

\section{Introduction} 


Large Language Models (LLMs), built upon the popular Transformer architecture and massive training datasets, are revolutionizing the way we interact with language. Their ability to ``understand'' complex text, generate responses, and translate between languages has far-reaching implications across industries.  Among these, the telecommunications domain stands to benefit significantly from the advancements in LLMs.

Telecommunications professionals are tasked with navigating the intricate and evolving landscape of standard specification documents established by bodies such as 3GPP, ETSI, and O-RAN. These standards define the fundamental technologies that enable global connectivity and interoperability.  AI-powered assistants based on LLMs have the potential to revolutionize how professionals interact with these standards. Such assistants could rapidly analyze dense specifications, generate code snippets, assist in debugging, and even offer insights that bridge the gap between different standards. This holds the promise of increased efficiency, innovation, and the streamlined development of robust telecom solutions.

However, conventional LLM-based assistants often struggle with knowledge-intensive tasks, a point discussed in detail in Section~\ref{sec:llms} of this paper. While LLMs can store factual information implicitly within their vast parameter sets, this knowledge is not easily accessible, updated, or traceable. In the context of telecommunications, where precision and reliance on standard-specified sources are critical, this limitation creates hurdles for widespread adoption in the industry.

To overcome these challenges, there is a growing need for specialized LLM-based solutions tailored to telecommunication standards. Section~\ref{sec:llms} explores various domains where LLMs have shown success, highlighting the need for such adaptation within telecommunications. Retrieval-Augmented Generation (RAG), a technique that combines LLM capabilities with a non-parametric knowledge base, offers a promising avenue for addressing the need for precise and factual answers.  Accordingly, we discuss in Section~\ref{sec:solution}  the challenges of applying this approach specifically to telecommunication standards, followed by the design of an architecture that tackles these challenges to build a Telecommunication Standards Assistant that is accurate, technically deep, and provides verifiable answers.

In Section~\ref{sec:implementation-and-evaluation}, we first present an implementation of this framework. Our work leverages state-of-the-art APIs and libraries to ingest telecom standards documents, create knowledge embeddings, and seamlessly integrate them within a user-friendly interface. We have evaluated our assistant with a vast pool of technical questions, one of which is presented in Section~\ref{sec:implementation-and-evaluation} as an example, rendering accurate, thorough, technically rich, and verifiable answers in marked contrast with the responses provided by general-purpose LLM-based assistants such as ChatGPT-4 or Gemini Ultra. 

\section{LLMs and Language Generation}\label{sec:llms}

This section provides some background on LLMs, their abilities (Sec.~\ref{sec:llm_intro}), and LLMs for specific application domains (Sec.~\ref{sec:llm_domains}). Then, we review the related work retrieval-augmented methods (Sec.~\ref{sec:rag_intro}), which are the base of our solution.

\subsection{Background on LLMs}\label{sec:llm_intro}
Large language models (LLMs) are Transformer-based language models with hundreds of billions or more parameters trained on an enormous corpus of text. Some examples include Gemini, LLaMa, and GPT-4. These models demonstrate significant capabilities in comprehending natural language and tackling intricate tasks through text generation. 
Although existing LLMs adopt the same Transformer architecture and pre-training objectives as other smaller language models, the significant increase in the model size, data size, and computing capacity largely improves the performance of LLMs. In fact, some works propose different scaling laws of LLMs, showing that model performance has strong dependencies with these tree dimensions \cite{scaling_law_1}. 

One of the most prominent features of LLMs is the emergent abilities, formally defined as the abilities that are not present in smaller models but arise in larger ones. 
The authors in \cite{emergent_abilities} show several examples of emergent abilities, where the performance in a task via few-shot prompting is random until a particular scale. When the model size surpasses a certain scale the performance increases well above random. This ability, also referred to as \textit{in-context learning}, is formally introduced by GPT-3 \cite{gpt3} but is not shown by previous models such as GPT-1 and GPT-2.
Another ability that emerged in large models is the step-by-step reasoning also called \textit{chain of thoughts} \cite{chain_thoughts}. This strategy is very effective in solving complex tasks that involve multiple steps such as math problems.

As emergent abilities cannot be predicted through the scaling laws of LLMs, this raises the question of whether additional scaling could potentially further expand the range of capabilities of language models \cite{emergent_abilities}.

\subsection{LLMs for specific application domains}\label{sec:llm_domains}
LLMs have been applied to different specific domain areas. In healthcare, Med-PaLM \cite{medpalm} achieves expert-level performance on the United States Medical Licensing Examination (USMLE). It relies on the pre-trained PaLM model enhanced with several techniques, namely, few-shot prompting, chain-of-thought and self-consistency prompting, and prompt tuning. However, the use of LLMs in healthcare comes with the risk of fabricating medical misinformation (e.g., suggesting advice inconsistent with medical guidelines) \cite{medical-misinformation}.

Finance is another important field where the application of LLMs is promising. In this field, BloombergGPT \cite{bloomberggpt} is a proprietary model based on BLOOM architecture trained with a mix of general purpose and specialized finance data. This results in a model that achieves SoTA results in financial benchmarks, while keeping competitive performance on general-purpose tasks.
As an alternative, FinGPT is an open-source framework comprising several layers for data gathering and engineering, pre-trained LLM models (e.g., LLaMA), and different fine-tuning techniques such as Fine-tuning via Low-rank Adaptation (LoRA) and Reinforcement Learning from Human Feedback (RLHF).

In the domain of scientific research, LLMs specially pre-trained on scientific-related corpora have been proposed (e.g. Galactica \cite{galactica}, Minerva \cite{minerva}). These models aim to assist in different stages of the scientific research pipeline.

However, all these solutions have the potential risk of hallucination, i.e., the generation of content that is not grounded in factual information but is instead a product of the model's creative extrapolation. This phenomenon occurs even when the models are trained with a high-quality and highly curated corpus of data, as in the case of Galactica \cite{galactica}. Moreover, hallucination may have serious implications in critical domains such as healthcare \cite{medical-misinformation}.
In order to reduce the hallucination phenomenon and provide reliable factual information, other  LLM-based approaches have been proposed as explained in the next section.

\subsection{Retrieval-Augmented Methods}\label{sec:rag_intro}

LLMs have demonstrated the ability to retain factual information within their parameters and excel in achieving state-of-the-art performance when fine-tuned for natural language processing tasks. 
However, the acquired world knowledge is implicitly encoded within the parameters of the underlying neural network, which poses a challenge in discerning what knowledge is stored in the network and where.
Consequently, LLMs have serious limitations to $i$) access and precisely manipulate knowledge; $ii$) add new information or include updates to the current knowledge. 
Thus, task-specific architectures can outperform language models in knowledge-intensive tasks\footnote{This term refers to tasks that humans could not reasonably be expected to accomplish without relying on an external knowledge source.} \cite{realm, rag}.

To improve the performance in knowledge-intensive tasks, some works in the literature propose hybrid models that combine a non-parametric memory (e.g., a textual knowledge corpus such as Wikipedia) with a pre-trained parametric memory (e.g., a LLM) \cite{realm, rag}. This architecture retrieves relevant information from the non-parametric memory, which is later processed and interpreted by the parametric memory (usually a language model), resulting in more factual and specific answers.
Moreover, this architecture allows us to directly modify or expand the knowledge stored in the parametric memory without having to modify the parametric memory.

REALM \cite{realm} augments a masked language model with a differentiable retriever, which gathers knowledge from a textual knowledge corpus. Both the language model and the retriever are trained together end-to-end according to the language model objectives. This strategy poses a significant computational challenge since the backpropagation through the retriever must consider millions of documents in the textual knowledge corpus. Moreover, REALM only explores open-domain extractive question answering, i.e., its objective is to predict masked tokens in a sentence. 

The authors in \cite{rag} expand this idea by proposing the retrieval-augmented generation (RAG) methodology for sequence-to-sequence (seq2seq) language generation. RAG considers a retriever based on Dense Passage Retrieval (DPR) \cite{dpr} and a seq2seq model (e.g., BART architecture) for language generation. The DPR retriever follows a bi-encoder architecture, one for the queries and the other for the documents in the knowledge corpus.
This method can be fine-tuned on any seq2seq task, whereby both the generator and retriever are jointly learned. In the fine-tuning phase, only the query encoder is trained. This reduces considerably the computational burden compared to REALM, which periodically updates the document index during pre-training \cite{realm}.
The results in \cite{rag} show that RAG responses are more factual, specific, and diverse than other baselines based on language models.
For these reasons and given the nature of the task involved in telecommunication standards, we rely on RAG to design TelecomRAG as detailed in the next section.

\section{TelecomRAG Design}\label{sec:solution}

\begin{figure*}[t]
    \centering
    \includegraphics[width=0.85\textwidth]{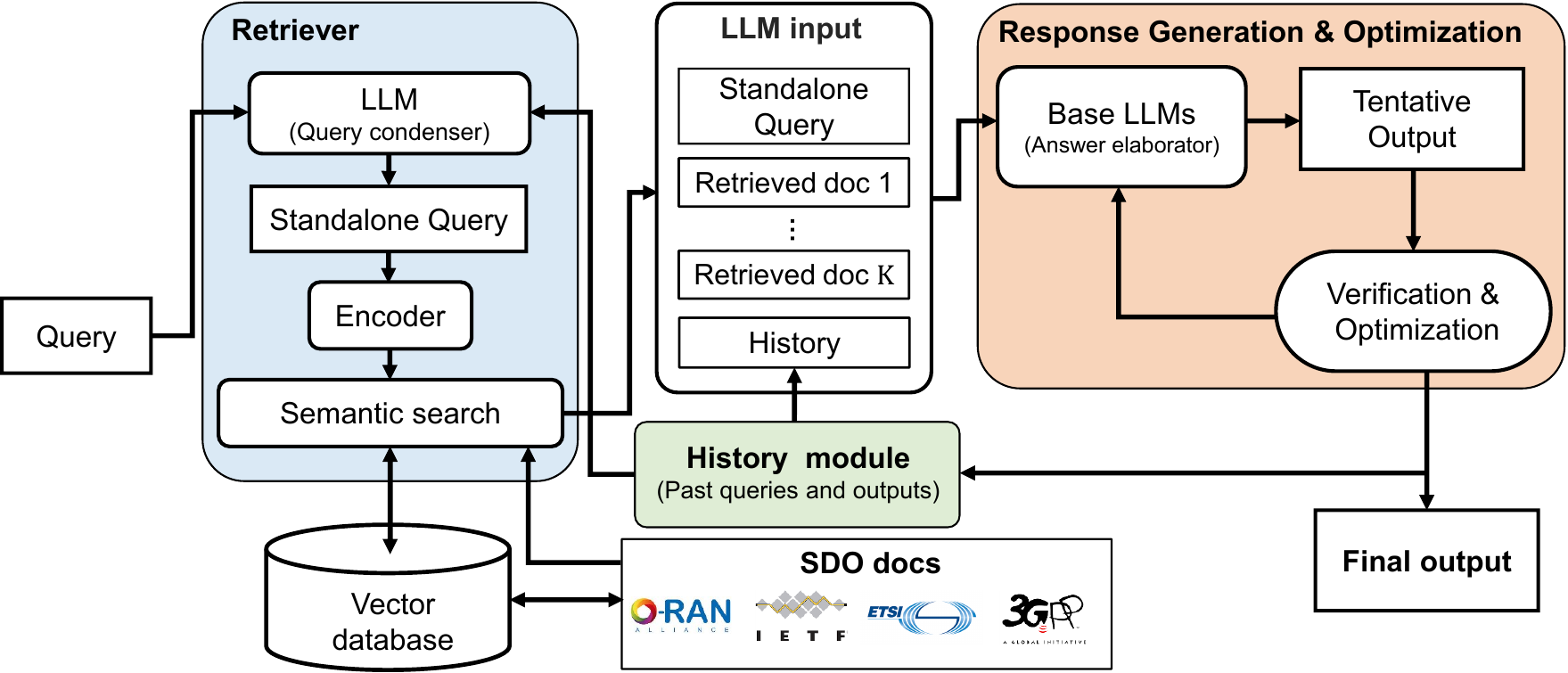}
    \caption{TelecomRAG: Architecture of the RAG-based Telecommunication Standards Assistant.}
    \label{fig:scheme}
\end{figure*}

This section details the conceptual design of our solution, which is mainly based on RAG with some practical modifications.
This solution combines information retrieval techniques and language models. As detailed in the previous section, this combination outperforms language models alone in knowledge-intensive tasks by providing more factual answers. Moreover, there are several advantages of generating answers instead of just extracting them directly from the knowledge base. For example, we may need to combine and elaborate the information included in several documents to obtain the answer to a specific query. Also, sometimes documents can contain clues about an answer but not the literal response.

TelecomRAG comprises two stages. In the first stage, we generate a vector database from the Telecom knowledge base. This stage is executed offline before launching the system. The vector database can be updated whenever new files need to be considered (e.g., new 3GPP releases or document updates).
The second stage is executed for every user query and establishes the conversational question-answering pipeline.

\subsection{Generation of Telecom vector database (Offline stage).}
The vector database creation involves the generation of embedding representations of the files in the Telecom knowledge base. 
Embeddings are dense vector representations in a latent space and they aim to capture the semantic meaning of the content. Thus, the embedding of two different documents should be close in the vectorial space if the content of these documents is related (even if they use completely different words). Based on this, documents with content related to a query can be retrieved and the system's output can be grounded on factual information contained in these documents.

As an initial step, the files in the Telecom Knowledge Base need to be cleaned and chunked. 
The configuration of the chunking is crucial. Overly short chunks provide insufficient information, while excessively long passages hinder the creation of a rich semantic space and may exceed the LLM's context window limit. The documents are divided into passages (referred to as \textit{documents} henceforth) of a fixed number of words with an overlap between consecutive chunks.
For each document, one embedding is computed and stored in the vector database. A common practice to compute the embedding is to use a language model based on the BERT architecture \cite{rag, bert}.

\subsection{Document Retrieval and Presentation (Online stage).}

The pipeline of the online stage is depicted in Fig.~\ref{fig:scheme}, comprising several building blocks:  $i$) the retriever, which processes the input query, gathers related documents from the knowledge base, and outputs the LLM input; $ii$) the history module, which stores past queries and responses; iii) the response generation and optimization module, which receives the LLM input and generates the final output. In the following, we detail the steps of the online pipeline.

First, the \textit{retriever} generates the \textit{standalone query} based on the user's new query and the history of past queries and responses. It is important the standalone query to be self-contained for the retrieval step. If we only use the new query, relevant context content may be missing. Conversely, if we input the full past conversation, irrelevant information may hinder an accurate retrieval. Therefore, the standalone query is generated by an LLM taking as input the new user's query and the history provided by the \textit{history module}.

Second, the embedding of the standalone query is computed using an encoder. We use the same encoder as in the offline stage allowing the documents and the queries to be encoded in the same latent semantic space. 

Third, we perform a semantic search in the vector database. We look for document embeddings that are similar to the encoded query based on a similarity metric. As mentioned before, the embeddings capture the ``gist'' of the text regardless of its length or the specific wording. Given that the dimensionality of the embeddings is usually high and the vector database can contain a large number of embeddings, an efficient search algorithm is needed to reduce the computational burden.
We retrieve $K$ documents from the knowledge base whose embeddings obtained the highest similarity metric value.

Fourth, we compose the \textit{LLM input} with the standalone query, the $K$ retrieved documents, and the history of queries and outputs, as shown in Fig.~\ref{fig:scheme}. The LLM input is passed to the Base LLM in the \textit{response generation \& optimization} module.

Fifth, the Base LLM generates a \textit{tentative output} based on the LLM input. To adapt the answer to different contexts, we use role-playing prompting techniques. In some cases, we may need a very detailed and comprehensive answer. To this end, we can assign the LLM the role of a ``standard expert'' and the user the role of a ``non-experienced trainee''. Based on this, we can generate a prompt as follows: \textit{``Assume you are a 3GPP standard expert and need to provide a very comprehensive answer to a non-experienced trainee.''} 
Thus, the tone and verbosity of the output can be easily adjusted.

Sixth, the tentative output generated by the Base LLM is checked by the \textit{verification \& optimization module}. The goal of this module is to check the appropriateness of the output. For example, if no documents are retrieved, a predefined message is sent to the user informing that there are no documents in the knowledge base related to the input query. Also, improper outputs can be filtered based on keywords. 
In other cases, the user can actively participate in the verification \& optimization loop by, for example, improving the query or discarding irrelevant parts of the retrieved documents or history manually.
Once the verification \& optimization is passed, the system outputs the final response.


\begin{figure}[b!]
    \centering
    \includegraphics[width=\columnwidth]{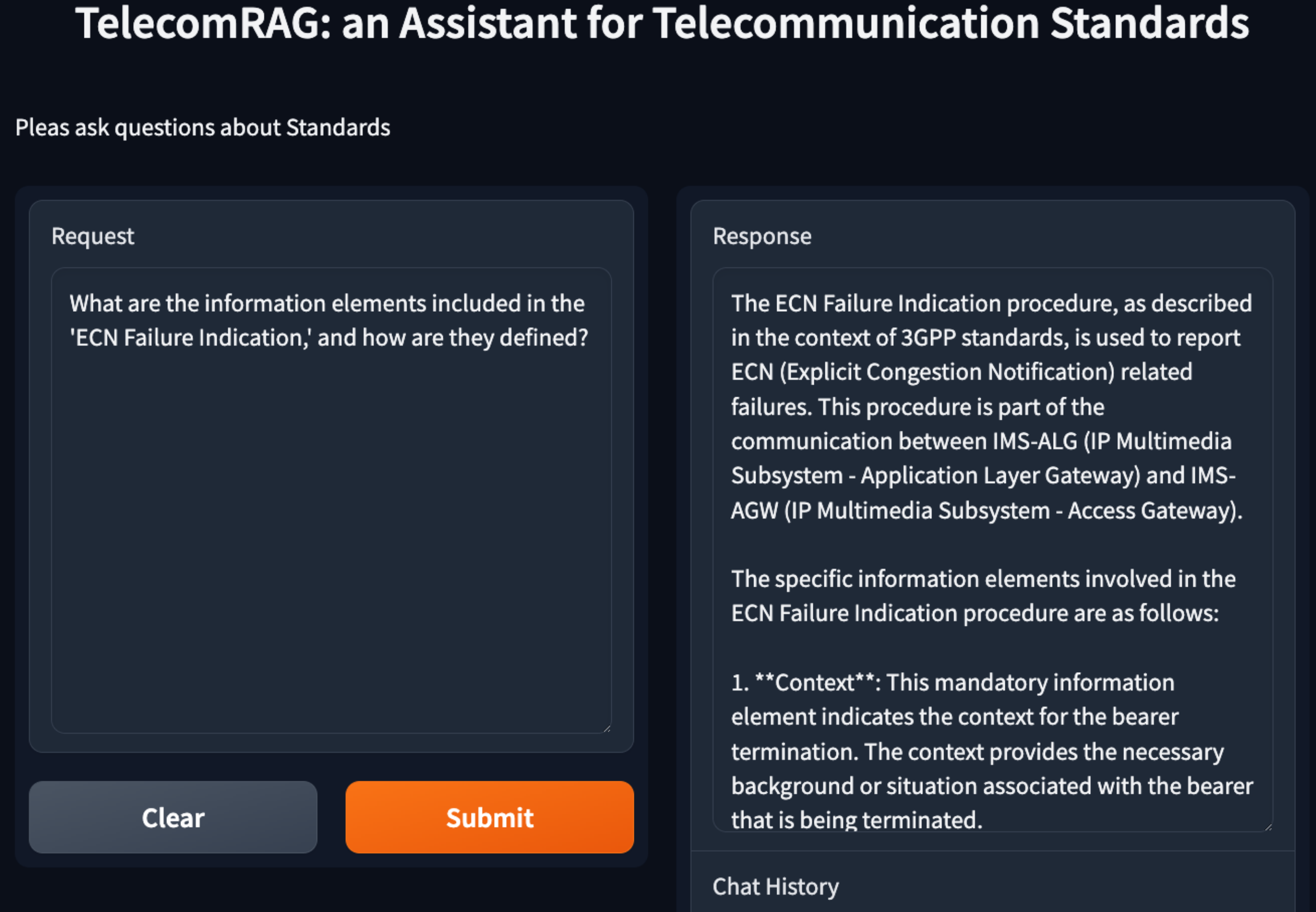}
    \caption{Graphical User Interface of TelecomRAG, an Assistant for Telecommunication Standards.}
    \label{fig:telecomllm-nle}
\end{figure}

\begin{table*}[t!]
\caption{ECN Failure Indication data from the correct 3GPP specification document~\cite[Section 8.30]{3GPP_TS_23334_2021}.}
\centering
\label{table:groundtruth}
\begin{tabular}{|l|l|l|l|p{8.5cm}|}
\hline
\textbf{Procedure} &
  \textbf{Initiated} &
  \textbf{Information element name} &
  \begin{tabular}[c]{@{}l@{}}\textbf{Information} \\ \textbf{element} \\ \textbf{required}\end{tabular} &
  \textbf{Information element description} \\ \hline
\multirow{3}{*}{\begin{tabular}[c]{@{}l@{}}ECN Failure \\ Indication\end{tabular}} &
  \multirow{3}{*}{IMS-AGW} &
  Context &
  Mandatory &
  This information element indicates the context for the bearer termination. \\ \cline{3-5} 
 &
   &
  Bearer Termination &
  Mandatory &
  This information element indicates the bearer termination for which the ECN  failure is reported. \\ \cline{3-5} 
 &
   &
  ECN Error Indication &
  Mandatory &
  This information element indicates an ECN failure event. \\ \hline
\multirow{2}{*}{\begin{tabular}[c]{@{}l@{}}ECN Failure\\Indication Ack\end{tabular}} &
  \multirow{2}{*}{IMS-ALG} &
  Context &
  Mandatory &
  This information element indicates the context where the command was executed. \\ \cline{3-5} 
 &
   &
  Bearer Termination &
  Mandatory &
  This information element indicates the bearer termination where the command was executed. \\ \hline
\end{tabular}
\end{table*}

\begin{table*}[t!] 
\caption{\textit{What are the information elements included in the ``ECN Failure Indication'', and how are they defined?}}
\footnotesize
\label{tab:evaluation} 
\begin{tabular}{c|p{17cm}>{\arraybackslash}m{18cm}} 
\hline
\rotatebox[origin=r]{90}{\textbf{TelecomRAG's response}} & 
The ECN Failure Indication procedure, as described in the context of 3GPP standards, is used to report ECN (Explicit Congestion Notification) related failures. This procedure is part of the communication between IMS-ALG (IP Multimedia Subsystem - Application Layer Gateway) and IMS-AGW (IP Multimedia Subsystem - Access Gateway).
 
The specific information elements involved in the ECN Failure Indication procedure are as follows:
\begin{enumerate}
\item \textbf{Context}: This mandatory information element indicates the context for the bearer termination. The context provides the necessary background or situation associated with the bearer that is being terminated. 
\item \textbf{Bearer Termination}: This mandatory information element indicates the bearer termination for which the ECN failure is being reported. The bearer termination refers to the point where a particular data session or communication is being terminated.
\item \textbf{ECN Error Indication}: This mandatory information element indicates an ECN failure event. An ECN failure event is a situation where the ECN mechanism has detected congestion in the network, but due to some error, the congestion notification could not be conveyed or processed as expected.
\end{enumerate}
In the ECN Failure Indication procedure, when the IMS-AGW receives any IP headers indicating ECN Congestion Experienced (ECN-CE), it should trigger rate adaptation in accordance with the specifications given in 3GPP TS 26.114. If there is an ECN failure, the IMS-AGW reports this failure to the IMS-ALG using the ECN Failure Indication message containing the aforementioned information elements.
 
Additionally, the ECN Failure Indication Acknowledgment (Ack) is used to acknowledge the receipt of the ECN Failure Indication. It also contains the Context and Bearer Termination information elements, which indicate the context and bearer termination where the command was executed.
\newline 

{\bf References:} ``IP Multimedia Subsystem (IMS) Application Level Gateway (IMS-ALG) - IMS Access Gateway (IMS-AGW) interface'' 3GPP Technical Specification 23.334
\\ \hline
\end{tabular}
\end{table*}

\section{TelecomRAG: Implementation and Evaluation}\label{sec:implementation-and-evaluation}

In the following, we present our implementation of the above design (Section \ref{sec:implementation}) and then, using an example query, we compare our implementation against vanilla chat assistants commonly used today (Section \ref{sec:evaluation}).

\subsection{Implementation}\label{sec:implementation}

Our implementation of the RAG-based Telecommunication Standard assistant design presented above is mostly founded on \texttt{Langchain}~\cite{topsakal2023creating}, a framework that provides an array of built-in wrappers and utilities designed to extract reasoning capabilities from language models. 

To build a knowledge base for telecommunication standards specialization, we gathered the complete set of 3GPP release-16 and release-18 standard specification documents. These documents are in various formats (PDF, TXT, DOCX, DOC), which can be readily processed by the \texttt{DirectoryLoader} method from the \texttt{Langchain.document\_loaders} library, along with custom functions for progress tracking, multi-threading, and error handling (e.g., corrupted file formats).

We divided the data into documents of 4000 characters with a 100-characters overlap between consecutive chunks using the \texttt{CharacterTextSplitter} method from \texttt{Langchain.text\_splitter}. For every document, we then generated an embedding using the OpenAI's \texttt{text-embedding-ada-002} embedding model via the \texttt{Langchain.embeddings.openai} library. This embedding model is also user to encode the standalone query at the retriever.
After all the text documents are encoded, we store them in a \emph{vector database} using \texttt{Chroma}, an AI-native vector database from \texttt{Langchain.vectorstores}.  

We implement the pipeline of the online stage using the \texttt{ConversationalRetrievalChain} method from \texttt{Langchain.chains} library. We use the LLM \texttt{gpt-4-1106-preview} from OpenAI for the query condensation (at the retriever) and for the answer elaboration (at the response generation \& optimization module).

We use \emph{cosine similarity} as the similarity metric in our system. Cosine similarity is a computationally cheap solution that has proven good performance in this type of application.
The semantic search is performed using an approximate nearest neighbor (ANN) algorithm called Hierarchical Navigable Small World (HNSW) implemented in \texttt{Chroma}.
We consider the $K=4$ documents with the highest cosine similarity obtained by HNSW algorithm to be included in the LLM input. 
The system's pipeline is conveniently wrapped up into a user-friendly web interface using \texttt{Gradio} that resembles that of conventional AI chatbots such as \texttt{Gemini} or \texttt{ChatGPT}, as shown in Fig.~\ref{fig:telecomllm-nle}.

\subsection{Evaluation}\label{sec:evaluation}

We evaluated TelecomRAG on a large set of technical questions derived from 3GPP specifications. The assistant consistently provided thorough, accurate, and technically detailed responses useful for telecom engineers, including references to specific 3GPP documents for verification.

For example, consider the query \textit{``What are the information elements included in the 'ECN Failure Indication', and how are they defined?''}, which is described in \cite{3GPP_TS_23334_2021} (Section 8.30) and summarized in Table~\ref{table:groundtruth} for validation. Table~\ref{tab:evaluation} depicts the response provided by our solution, including a reference to the right 3GPP document, demonstrating  accuraccy and verifiability for this example. Notably, our solution maintained this depth and accuracy across all queries in our evaluation. 

In marked contrast, when issuing the same queries to general-purpose assistants such as ChatGPT-4, Gemini Ultra or TelecomGPT (a third-party application built on ChatGPT to provide specialization on the telecommunication sector\footnote{Available in \url{https://chat.openai.com/g/g-eSwyzwwDV-telecom-gpt}.}), the responses provided are vague, inaccurate and not verifiable, as illustrated in Table~\ref{tab:comparison} for the query \textit{``What are the information elements included in the 'ECN Failure Indication', and how are they defined?''} presented earlier.

\begin{table*} 
\caption{\textit{What are the information elements included in the ``ECN Failure Indication'', and how are they defined?}}
\footnotesize
\label{tab:comparison} 
\begin{tabular}{c|p{17cm}>{\arraybackslash}m{18cm}} 
\hline
\rotatebox[origin=r]{90}{\textbf{ChatGPT 4's response}} & The ``ECN Failure Indication'' is a term that is typically associated with network protocols, specifically with mechanisms that deal with congestion control. ECN stands for Explicit Congestion Notification. This feature enables network nodes to notify each other about network congestion without dropping packets. However, the specific term ``ECN Failure Indication'' isn't standard, and you may be referring to situations where ECN signaling fails or is not properly communicated between network entities. Here's a detailed look at the typical elements related to ECN, which might be relevant to a failure scenario:
\begin{enumerate}
\item	\textbf{ECN-Capable Transport (ECT)} - This is a codepoint in the IP header that indicates that the end-points of the TCP connection are ECN-capable, meaning they can both send and receive ECN signals. There are two ECT codepoints, ECT(0) and ECT(1), which serve slightly different purposes but essentially signal ECN capabilities.
\item	\textbf{Congestion Experienced (CE)} - This is another codepoint in the IP header used by routers experiencing congestion to indicate this status to the end-points of the transmission. When a packet marked with an ECT codepoint encounters congestion, routers that support ECN can mark these packets as CE instead of dropping them.
\item	\textbf{ECN Echo (ECE)} - This is a flag in the TCP header used to indicate that a segment with the CE codepoint has been received. The receiving end uses this flag to notify the sender about the congestion.
\item	\textbf{Congestion Window Reduced (CWR)} - This flag is set by the sending end to signal to the receiving end that it has received the ECE flag and has reduced its congestion window size accordingly. This is an acknowledgment that the sender is responding to the congestion.
\end{enumerate}
When discussing ``ECN Failure Indication'', it might relate to any anomalies or failures in the signaling process among these elements. Here are a few scenarios where ECN could ``fail'':
\begin{itemize}
\item	\textbf{Unsupported ECN:} If any part of the path between sender and receiver does not support ECN, then the ECN signals (ECT, CE) may not be handled or propagated correctly.
\item	\textbf{Incorrect ECN Implementation:} Bugs or incorrect implementations in routers or end hosts can lead to incorrect ECN behavior, such as failing to set CE flags despite congestion or not reacting to ECE flags.
\item	\textbf{ECN Negotiation Failure:} During the TCP handshake, both sides must agree to use ECN. If there is a failure in this negotiation, ECN will not be used, even if both endpoints support it.
\end{itemize}
Understanding specific ECN failure modes would depend on the context in which ECN is being used (like in specific network architectures or for certain types of traffic) and the specific tools and methods used for diagnosing such failures.  
\\ \hline
\rotatebox[origin=r]{90}{\textbf{Gemini Ultra's response}} & 
Unfortunately, there isn't a universally defined ``ECN Failure Indication'' with a set list of information elements.  ``ECN'' can refer to different things depending on the context, and the specific details of a failure indication would depend on the particular system or protocol involved.
Here's a breakdown of the possibilities:
\begin{itemize}
\item \textbf{IP Networking:} In the context of Internet Protocol (IP) networking, ECN (Explicit Congestion Notification) is a mechanism for routers to signal congestion to senders. It's unlikely there would be a separate ``ECN Failure Indication''. Issues with ECN functionality might be reported through other mechanisms within the specific routing protocol.
\item	\textbf{Other Fields:} ECN can also stand for Engineering Change Notice or Entry Change Notice in various fields. A failure related to an ECN would likely be specific to the system or process that uses those terms and wouldn't have a standardized format.
\end{itemize}
Here are some ways to get more information:
\begin{itemize}
\item	\textbf{Context Clues:} If you encountered the term ``ECN Failure Indication'' somewhere specific, look for surrounding information that might indicate the context (e.g., manuals, software documentation).
\item	\textbf{System Identification:} If you know the system or protocol where you came across this term, searching for its documentation might reveal details about error reporting mechanisms.
\end{itemize}
If you can provide more context about where you encountered ``ECN Failure Indication'', I might be able to offer more specific guidance.
\\ \hline
\rotatebox[origin=r]{90}{\textbf{TelecomGPT response}} & 
The ``ECN Failure Indication'' message in the context of telecommunications and protocols isn't explicitly defined in the documents I have available, such as the SMPP specifications and various telecom-related white papers and presentations. It appears you might be referring to a specific protocol message or error indication that might be part of a broader telecommunications standard or proprietary system not covered in these documents.
If the ``ECN Failure Indication'' you are asking about is related to a specific application, protocol, or vendor-specific implementation, it would be helpful to have more context or details about the system in which this term is used. If it is part of a standard not covered in the documents or discussions I have access to, you may need to consult the specific documentation or standards applicable to that protocol or system. If you can provide more details or clarify the specific protocol or context in which you encountered ``ECN Failure Indication'', I can assist you better. Alternatively, if this is related to a standard or system, the relevant documentation or technical support resources for that system may be the best source for detailed information.
\\ \hline
\end{tabular}
\end{table*}

\section{Conclusions} 

This paper presented a specialized LLM solution tailored to telecommunication standards. This solution is based on RAG and combines information retrieval techniques and LLMs, providing precise, fact-based, and verifiable answers. We evaluate our solution using a knowledge base built from 3GPP Release 18 specification documents. Our solution surpasses generic LLMs, offering superior accuracy, technical depth, and verifiability. In the next phase, we plan to extend the knowledge base with other SDO documentation.
We believe that this work can bring significant value to the telecommunications field. 

\bibliographystyle{IEEEtran}
\bibliography{references}

\begin{IEEEbiographynophoto}{Girma M. Yilma} 
 is a Senior Research Engineer at NEC Laboratories Europe in Heidelberg, Germany. He received his M.Sc. in Telecommunications Engineering from the University of Trento, Italy. His current research interests are focused on GenAI for Telecom Services, Trusted Execution Environments for NFV MANO, and Cloud Native Networking.
\end{IEEEbiographynophoto}

\begin{IEEEbiographynophoto}{Jose A. Ayala-Romero} received his Ph.D. degree from the Technical University of Cartagena, Spain, in 2019. Then, he joined Trinity College Dublin, Ireland, as a post-doctoral researcher. From January 2021 to September 2022, he was a Senior Data Scientist at Huawei Ireland Research Center, Ireland. Currently, he is a Senior Researcher at NEC Laboratories Europe, Germany. His research interests include the application of diverse machine learning techniques to mobile networks problems.
\end{IEEEbiographynophoto}

\begin{IEEEbiographynophoto}{Andres Garcia-Saavedra} received his PhD degree from the University Carlos III of Madrid (UC3M) in 2013. He then joined Trinity College Dublin (TCD), Ireland, as a research fellow until 2015.
Currently, he is a Principal Researcher at NEC Laboratories Europe. 
His research interests lie in the application of fundamental mathematics to real-life wireless communication systems.
\end{IEEEbiographynophoto}

\begin{IEEEbiographynophoto}{Xavier Costa-Perez} is a Research Professor at ICREA, Scientific Director at i2Cat and Head of 5G/6G R\&D at NEC Laboratories Europe. His team generates research results which are regularly published at top scientific venues, produces innovations which have received several awards for successful technology transfers, participates in major European Commission R\&D collaborative projects and contributes to standardization bodies. He received both his M.Sc. and Ph.D. degrees in Telecommunications from the Polytechnic University of Catalonia (UPC). 
\end{IEEEbiographynophoto}

\end{document}